# Plasmonic interferences sampled by a diffraction grating generate moiré fringes

*Marin Tharrault[1], Guillaume Boulliard[1], Emmanuel Lhuillier[2] and Aloyse Degiron[1,*]*

[1]Université Paris Cité, CNRS, Laboratoire Matériaux et Phénomènes Quantiques, 75013 Paris, France

[2]Sorbonne Université, CNRS, Institut des NanoSciences de Paris, INSP, F-75005 Paris, France

*aloyse.degiron@u-paris.fr

ABSTRACT: We report on the formation of moiré patterns when observing the diffraction of surface plasmons by periodic gratings of finite extent with an imaging spectrometer that maps the light emission as a function of the wavelength and the propagation distance. This phenomenon results from the formation of standing waves upon the partial reflection of surface plasmons at the far end of the grating and their scattering by the periodic corrugations. The moiré fringes are created as a result of the incommensurability between the sampling pitch of the grating and the wavelength-dependent standing waves. We introduce a scalar model that supports these observations and provides physical insight into these results, including a method to estimate the complex Fresnel reflection coefficient of a surface plasmon at the edge of a grating.

## 1. Introduction

Metallic gratings and surface plasmons have an intertwined history, both being key ingredients in landmark studies such as the observation of Rayleigh-Wood anomalies [1], the shaping of fluorophore emission [2] or the discovery of the extraordinary optical transmission [3]. The reason for this interdependence is that a grating carved at the surface of a metal allows surface plasmons to couple with free space radiation by diffraction [4].

Metallic gratings can also be combined to form more complex structures such as moiré patterns. Moiré patterns are typically obtained by superimposing gratings having the same orientation but unequal periodicities, or by rotating two or more gratings with respect to one another. The countless resulting combinations (moiré super-lattices, moiré quasi-crystals, moiré cavities…) offer a rich variety of tools to manipulate surface plasmons [5–7], lower their group velocity [8–10], and use them in lasing schemes [11]. Equally interesting are hybrid moiré patterns involving surface plasmons and periodic luminescent patterns, with applications ranging from sub-wavelength imaging [12] to circularly polarized luminescence [13].

In this study, we report on plasmonic moiré patterns that are not physically imprinted in the sample, contrarily to the above works. Instead, they arise from the incommensurability between a plasmonic standing wave and the fixed periodicity of a diffraction grating. We observe the moiré fringes by analyzing the leakage of the surface plasmons propagating along the grating using an imaging spectrometer that provides spatial information on the plasmon propagation along one dimension and wavelength information along the other dimension. We support our findings with an analytical model that reproduces the salient features of the experiments. We finally discuss how non-trivial insight into the plasmon propagation can be extracted from the data.

**2. Experiments**

Figure 1(a) provides a schematic of the system under investigation as well as the principle of the experiments. The sample is a 200 nm thick Au layer decorated with a linear corrugated grating of finite spatial extent $D = 80$ µm. The height of the corrugations, which are fabricated by electron-beam lithography and Au deposition, is 50 nm, their periodicity $P$ is 1.35 µm and filling factor 25% [Fig. 1(b)]. The sample is subsequently coated with a thin layer of PbS nanocrystals (NCs) emitting in the near-infrared between 1.15 and 1.6 µm, as shown in Fig. 1(c) [14]. The layer of PbS NCs (between 15 to 20 nm thick, corresponding to approximately 3 monolayers of PbS NCs) does not play a role in the moiré phenomenon that is reported below. Its sole purpose is to give us a way to create a local and broadband source of surface plasmons, by focusing the light from a 633 nm HeNe laser onto the surface of the sample at a distance $L = 100$ µm away from the linear grating, in a flat area of the sample. This optical pumping in the visible absorption tail of the PbS NCs [15] makes them emit infrared light by photoluminescence (PL). Due to the subwavelength size of the emitters and their close proximity to the metal surface, this infrared PL preferentially occurs under the form of propagating surface plasmons. As elaborated elsewhere [16,17], the plasmons launched under this excitation scheme can be approximated as cylindrical surface waves that propagate away from the laser excitation area.

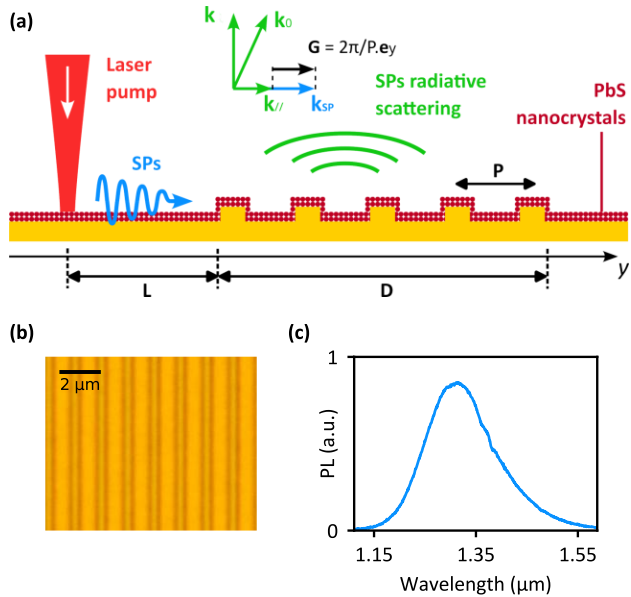

FIG 1. (a) Experimental configuration. The 632 nm HeNe laser pumps the PbS nanocrystals film, decaying into surface plasmons that propagate over a distance L = 100 µm before reaching a Au linear grating (of extent $D = 80$ µm and pitch $P = 1.35$ µm), where they are diffracted in free space following Eq. (1). (b) Optical microscope image of the grating under white light illumination. (c) Photo-luminescence of the NC film on flat Au

In our experiments, we analyze the light emitted by the surface plasmons as they reach and propagate along the metallic grating. To this end, we use the same 10X microscope objective to launch the surface plasmons outside the grating area and to collect the near-infrared light diffracted by the periodic linear corrugations. Figure 2(a) provides a direct visualization of the broadband emitted light. On top of the image, a very bright spot centered on $y = -L$ is visible that corresponds to the area where the HeNe laser is focused on the surface. The near-infrared PL at the location of this spot is so intense compared to the other features of the image that we display it with a saturated scale. Beyond this spot, no emission is apparent until we reach the edge of the grating. Light is suddenly emitted from there on and over the full extension of the grating, with decreasing intensity as the distance from the excitation spot increases.

These features are consistent with the radiative leakage of surface plasmons originating from the laser excitation spot. These plasmons first propagate along the flat surface of the sample with negligible coupling to free space. Once the plasmons reach the grating at $y = 0$, they become radiative due to scattering /diffraction. If the plasmons have not totally decayed through photon emission and Joule losses when they reach the outer end of the grating at $y = D$, they will essentially become non-radiative again as they propagate away from the grating. As for the PL intensity just under the laser excitation spot, it corresponds to the fraction of PbS NCs that emit light in free space rather than surface plasmons. Only a fraction of the initial energy reaches the edge of the grating where the plasmons become radiative.

To confirm this interpretation, we next analyze the properties of the light radiated from the sample. We recall that surface plasmons propagating on periodic gratings are diffracted into free space according to the momentum conservation law [4]:

$$\boldsymbol{k}_{//} = \boldsymbol{k}_{SP} \pm \boldsymbol{G} \qquad (1)$$

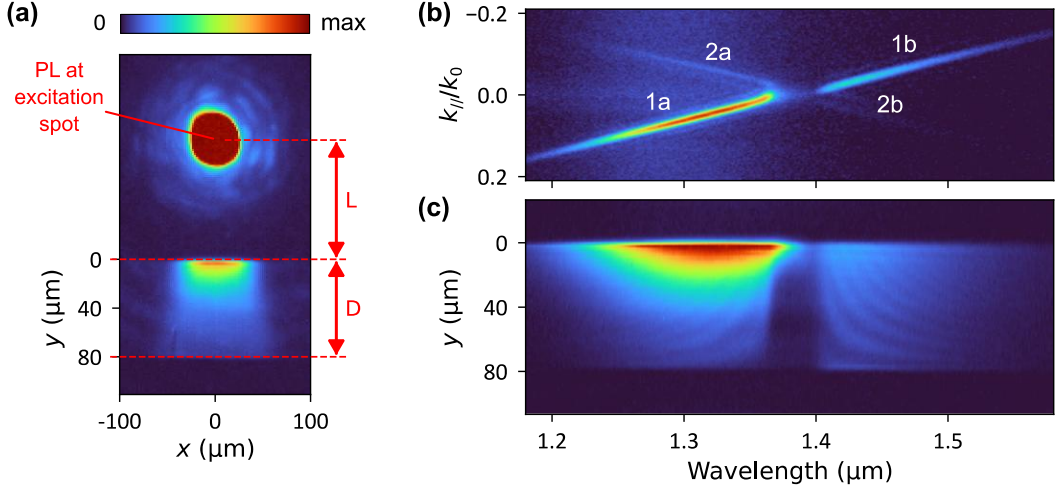

FIG. 2. (a) Real-space photoluminescence image. The surface plasmons are launched at a distance $L = 100$ µm away from the linear grating of extent D = 80 µm. The image is displayed with a saturated color scale to highlight the fine features of the measurements. (b) Experimental dispersion relation of the plasmons that are scattered in free space by the grating [the horizontal axis is common to panel (c) and represents the wavelength λ]. The forward surface plasmon branches are labelled 1a and 1b while the back-propagating ones are labelled 2a and 2b. (c) Spectral evolution of the same diffracted plasmons as a function of *y*. The color bar applies to all panels.

where $k_{SP}$ and $k_{//}$ are the plasmon and in-plane component of the free-space wavevectors, respectively, while $G$ is the reciprocal vector $G = 2\pi/P.\mathbf{e_y}$ of the grating [Fig. 1(a)]. This equation indicates that forward- ($k_{SP}>0$) and backward- ($k_{SP}<0$) propagating plasmons are shifted in the wavevector space by $\mp G$ and thus brought into the light cone. The experimental dispersion relation of the light emitted by the sample [14], plotted in Fig. 2(b), features the branches predicted by Eq. (1) [note that in this plot, the quantities are normalized by the norm of the free space wavevector $k_0$ and that Eq. (1) can be rewritten as $k_{//}/k_0 = k_{SP}/k_0 \pm \lambda/P$]. The bright branches (labelled 1a and 1b) correspond to the diffraction of forward-propagating plasmons, while the fainter branches (labelled as 2a and 2b) correspond to the diffraction of backward-propagating plasmons. The gap that is apparent in the spectral range [1.37 – 1.41 um] is a second-order effect that is well understood [18] and that will not be commented here. The reason why branches 1a and 1b are much brighter than branches 2a and 2b can be understood by the excitation conditions of our experiment: the plasmons are launched at a position y=-L from the grating and propagate in the forward $\mathbf{e_y}$ direction, solely contributing to the forward-propagating plasmon branches 1a and 1b. The existence of fainter branches 2a and 2b implies that some plasmons are back-reflected at the outer edge $y = D$ of the grating. This reflection arises due to the impedance mismatch between the grating and the flat film beyond it and plays a fundamental role in the phenomenon reported in this article.

We next study the spectrum of the emitted light as the plasmons propagate along the $y$ axis. To this end, we spatially filter a narrow vertical section of Fig. 2(a) that crosses the $x = 0$ abscissa and experimentally disperse the data as a function of the wavelength using an imaging spectrometer [14]. The results, displayed in Fig. 2(c), are consistent with the previous experimental observations: at all wavelengths, the light intensity is maximum at the edge of the grating that is closest to the laser excitation spot and then gradually decreases due to radiative and Joule losses and the fact that the plasmons spread as cylindrical waves. Furthermore, the spectral bandwidth matches that of the NC photoluminescence plotted in Fig. 1(c). Last, no light is emitted between 1.37 µm and 1.41 µm, which corresponds to the gap in the dispersion relation of Fig. 2(b) that indicates that plasmon propagation is forbidden at these wavelengths. But the most striking feature of this experiment is that the intensity is modulated by a moiré pattern along the propagation direction—resolved here as a function of the wavelength λ.

## **3. Model**

To understand the formation of this moiré pattern, we develop a scalar model in which the surface plasmons are described as cylindrical surface waves originating from the laser excitation spot, with an amplitude proportional to the square root of the experimental PL spectrum intensity $S(\lambda)$ of the layer of PbS NCs plotted in Fig. 1(c). Furthermore, we suppose that the presence of the grating between $y = 0$ and $y = D$ affects the waves in only two ways: (i) it induces partial reflections at both ends due to an impedance mismatch between the corrugated areas and the flat interfaces at $y = 0$ and $y = D$, (ii) it acts upon the complex effective index $n_{SP}(\lambda) = k_{SP}/k_0$—the real and imaginary values of which can be extracted from dispersion measurements [14]. From these measurements, plotted in Fig. S2, we notice that the plasmon propagation length is smaller than twice the spatial extension $D$ of the grating. Therefore, we can neglect multiple reflections between the two edges of the grating. The expression of the scalar field $A(\lambda, y)$ for $0 \leq y \leq D$ therefore consists of a forward-propagating cylindrical wave and a partially-reflected wave at $y = D$ with reflection coefficient $r$, weighted by $S(\lambda)$:

$$A(\lambda, y) \approx \sqrt{\frac{S(\lambda)}{y+L}} \exp[in_{SP}(\lambda)k_0(y+L)] + r\sqrt{\frac{S(\lambda)}{2D+L-y}} \exp[in_{SP}(\lambda)k_0(2D+L-y)] \qquad (2)$$

Eq. (2) has the characteristics of a standing wave pattern, with nodes periodically occurring to the tune of $\lambda/2$ (see schematic representation in Fig. S3).

We model the scattering of the field by the grating using a function $F(y)$, of same periodicity P as the grating. We assume it to be identical for both forward and reflected waves, so that the scattered field can be written as $F(y)A(\lambda, y)$. The light intensity collected into the imaging spectrometer thus reads as:

$$I(\lambda, y) \sim |F(y)A(\lambda, y)|^2 \qquad (3)$$

The scattering function $F(y)$ is a central part of our model. Had we supposed that light were emitted uniformly by the grating, it would have implied that our experimental apparatus creates an image of $|A(\lambda, y)|^2$ in which no moiré pattern is present (see Fig. S4).

We empirically found that the actual implementation of $F(y)$ is not critical to produce moiré fringes [19]. The fundamental reason for this freedom of choice is that the spatial details of $F(y)$ are below the diffraction limit and therefore not resolved by our experimental apparatus. Here we use a $P$-periodic Dirac comb:

$$F(y) = \sum_{m=0}^{\frac{D}{P}} \delta(y - mP) \qquad (4)$$

where $\delta$ is the Dirac distribution. A more thorough derivation, presented in the Appendix, shows that the moiré fringes are spaced along the $y$-axis by:

$$\frac{1}{2} P \left| \frac{Re[n_{SP}(\lambda)]P}{\lambda} - 1 \right|^{-1} \qquad (5)$$

As $\lambda$ approaches the plasmonic edges of the gap ($\lambda = 1.37$ µm and $\lambda = 1.41$ µm), Eq. (5) diverges, explaining the characteristic hyperbolic shape of the moiré fringes experimentally observed in Fig. 2(c). A visual count on this figure reveals that there are approximately eight fringes at $\lambda = 1.48$ µm (0.1 µm away from the grating bandgap), while Eq. (5) predicts a periodicity of 10 µm for this 80 µm-long grating, giving a total of 8 periods as well.

We now show that this model successfully captures the experimental observations by studying the evolution of the moiré pattern for three grating extensions $D = 40$, 80 and 120 µm. We numerically evaluate Eqs. (2-4) on a $\lambda - y$ grid. We introduce finite resolutions for both the optical system and the grating spectrometer, allowing to filter $I(\lambda, y)$ by a two dimensional point spread function (PSF) that takes into account the nominal spectral resolution of our spectrometer ($\sigma_\lambda = 2.6$ nm) and the theoretical [20] spatial resolution of our 10X objective with a numerical aperture of 0.3 ($\sigma_y \approx 1.0$ µm).

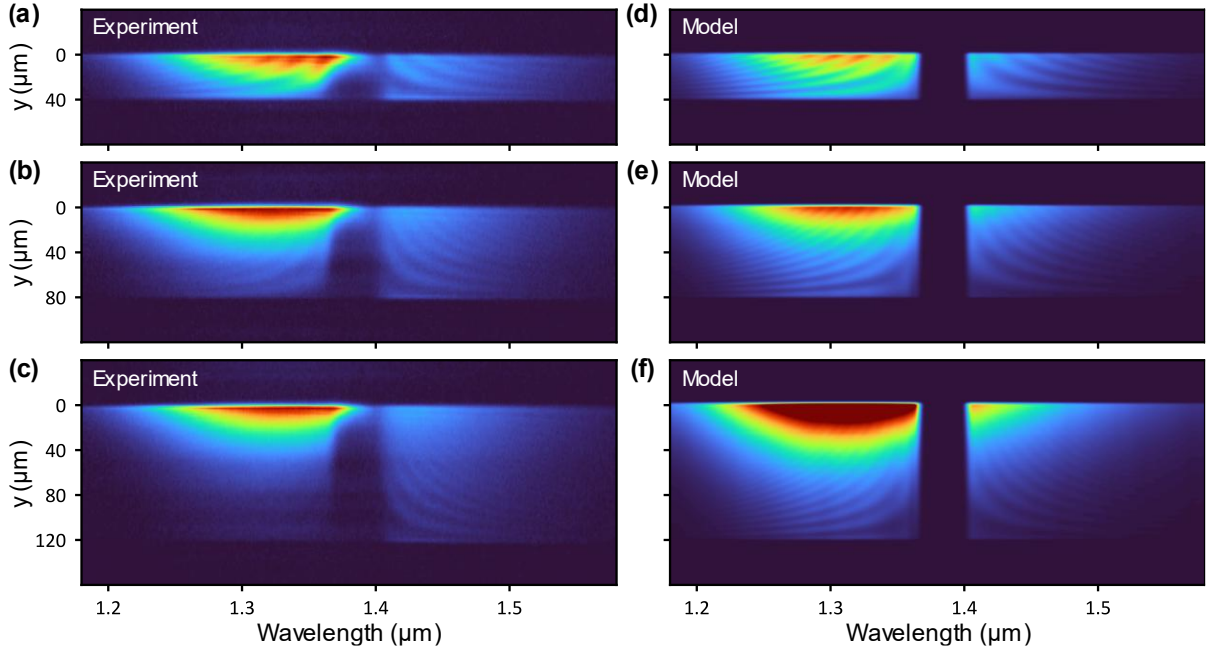

FIG. 3. (a-c) Experimental and (d-f) modeled spectral evolution of the plasmon scattering in free space as a function of y. Three grating lengths D are studied: (a, d) D = 40 µm, (b, e) 80 µm and (c, f) 120 µm. Same color code as Fig. 2.

The overall agreement between experiments [Fig. 3(a-c)] and calculations [Fig. 3(d-f)] is very good. Both sets of data show moiré patterns that gradually disappear as D increases, with fringes barely visible close to *y* = 0 for the longest extension. This result is a direct consequence of the attenuation of the surface plasmons, the reflected wave becoming too weak with respect to the forward wave to produce a significant contrast in the moiré fringes.

## 4. Discussion

While such moiré patterns are above all academic curiosities, they nevertheless offer valuable insights into the emission of light from a plasmonic grating. In particular, they provide an experimental means to probe and visualize, albeit indirectly, the existence of plasmonic interferences which would have otherwise gone unnoticed with a simple visualization of the broadband light emission [cf. Fig. 2(a)]. The formation of these standing waves is a manifestation of the extended spatial coherence of the plasmon modes, which reaches approximately 80 µm according to our estimations [Fig. S2], even though they have been excited with the incoherent spontaneous emission of colloidal NCs. As a corollary, this study shows that gratings with groove heights of only 50 nm, which is more than 20 times smaller than the wavelengths under investigation, are sufficiently perturbative to induce significant reflection between the corrugated and non-corrugated areas of the metallic film—an observation that can be rationalized by the fact that the evanescent field of a non-radiating surface plasmon propagating along a smooth surface has a poor overlap with the oscillatory field of a leaky one. Moreover, the moiré patterns allow one to make a rough estimation of the magnitude and phase of the partially-reflected waves: while the model produces moiré fringes for a broad range of reflection coefficients $r$ (Fig. S5), we found that using a reflection coefficient close or equal to $r = 0.2\exp(i\pi/2)$ provides the best agreement with the observations. This is the case shown in Fig. 3(d-f). Another interesting aspect is that the moiré patterns are clearly visible even though our experiment does not spatially resolve the periodic leakage $F(y)$ of the plasmons. Since this periodic leakage is key to the formation of the moiré fringes, their observation provides spatial information at scales that are not directly accessible experimentally. In this respect, our experiments are reminiscent of other works in which moiré patterns are exploited to probe features that are smaller than the resolution of the apparatus [12,21].

**Appendix: Analytical derivation of the moire periodicity**

We derive a simple model based on the understanding developed in the main text to extract the moiré oscillations and periodicity. For simplicity, the envelope profile due to the NC-mediated plasmon injection as well as the attenuation of the fields are ignored here.

For this computation we discriminate both forward and backwards components of the scalar field, $A_1$ and $A_2$:

$$A_{1,2} \sim \exp(\pm i n k_0 y) \tag{1}$$

That allows us to write the scattered fields $\widetilde{A_{1,2}}$ using the scattering function $F(y) = \sum_m \delta(y - mP)$, where we convolve the Dirac distribution with Gaussian distributions of width $\sigma$ to encompass the effect of the imaging system:

$$\widetilde{A_{1,2}} \sim [A_{1,2} \cdot F](y) * \exp\left(-\frac{y^2}{2\sigma^2}\right) \sim \sum_m \exp(\pm i n k_0 mP) \times \exp\left[-\frac{(y-mP)^2}{2\sigma^2}\right] \tag{A2}$$

where $*$ is the convolution product. We then operate the substitution $l = m - \lfloor y/P \rfloor$ (where $\lfloor x \rfloor$ is the floor function of $x$), which, given that the resolution of the imaging system is much wider than the periodicity $P$, allows us to consider that the exponential term is equal to one up to a cutoff $|l| < \sigma/P$ and zero otherwise.

$$\widetilde{A_{1,2}} \sim \sum_{|l| < \sigma/P} \exp\left[\pm i n k_0 \left(l + \left\lfloor \frac{y}{P} \right\rfloor\right) P\right] \tag{A3}$$

Note that the identical scattering mechanism introduced here for both the forward and reflected waves results in the same position-dependent term $\lfloor y/P \rfloor$. The more general case would have produced two slightly different terms.

The moiré pattern arises from the beat between these two waves:

$$\left|\widetilde{A_1} + \widetilde{A_2}\right|^2 \sim \left|\widetilde{A_1}\right|^2 + \left|\widetilde{A_2}\right|^2 + 2\Re\{\widetilde{A_1} \cdot \widetilde{A_2}^*\} \tag{A4}$$

Computing the product in the last term we obtain:

$$\widetilde{A_1}.\widetilde{A_2}^* = \exp\left(2ink_0 \left\lfloor \frac{y}{P}\right\rfloor P\right) \sum_{|l|,|l'|<\sigma/P} \exp[ink_0(l+l')P] \quad (A5)$$

The first position-dependent term encompasses the slowly-varying envelope of the moiré pattern. The wavelength being only slightly detuned from the grating periodicity, we can define a small phase parameter $|\varphi| \ll 1$ such that $nk_0 P = 2\pi - \varphi$. The first term can thus be rewritten as:

$$\exp\left(2ink_0 \left\lfloor \frac{y}{P}\right\rfloor P\right) = \exp\left[2i(2\pi - \varphi)\left\lfloor \frac{y}{P}\right\rfloor\right] = \exp\left(-2i\varphi \left\lfloor \frac{y}{P}\right\rfloor\right) \sim \exp(-2i\varphi y/P) \quad (A6)$$

Substituting $\varphi$ by its value and using $k_0 = \frac{2\pi}{\lambda}$, we obtain:

$$\widetilde{A_1}.\widetilde{A_2}^* \sim \exp\left(-2i\varphi \frac{y}{P}\right) = \exp\left(2i(nk_0 P - 2\pi)\frac{y}{P}\right) = \exp\left[2\pi i \times 2\left(\frac{nP}{\lambda} - 1\right)\frac{y}{P}\right] \quad (A7)$$

This last equation yields the moiré periodicity in position:

$$P_y = \frac{1}{2}\left|\frac{P}{nP/\lambda - 1}\right| \quad (A8)$$

Introducing a complex dielectric index $n$ would have left the analytical derivation unchanged except for exponentially decaying terms that do not alter the phase factor extracted here.

**Acknowledgements**

We acknowledge support from the European Research Council Grants FORWARD (grant n°771688) and AQDtive (grant n°101086358) as well as from French National Research Agency (ANR) through the grant Bright (ANR-21-CE24-0012-02). We acknowledge the use of clean-room facilities from the "Groupement des Plateformes de Micro et Nanotechnologies de Paris" and support from CNRS, Université Paris Cité and Renatech+ for micro and nanofabrication.

**References**

[1]     U. Fano, The Theory of Anomalous Diffraction Gratings and of Quasi-Stationary Waves on Metallic Surfaces (Sommerfeld's Waves), J. Opt. Soc. Am. **31**, 213 (1941).

[2]     W. Knoll, M. R. Philpott, J. D. Swalen, and A. Girlando, Emission of light from Ag metal gratings coated with dye monolayer assemblies, J. Chem. Phys. **75**, 4795 (1981).

[3]     T. W. Ebbesen, H. J. Lezec, H. F. Ghaemi, T. Thio, and P. A. Wolff, Extraordinary optical transmission through sub-wavelength hole arrays, Nature **391**, 667 (1998).

[4]     H. Raether, *Surface Plasmons on Smooth and Rough Surfaces and on Gratings*, Vol. 111 (Springer Berlin Heidelberg, 1988).

[5]     Z. Liu, S. Durant, H. Lee, Y. Xiong, Y. Pikus, C. Sun, and X. Zhang, Near-field Moiré effect mediated by surface plasmon polariton excitation, Opt. Lett. **32**, 629 (2007).

[6]     Z. Guo, Z. Y. Zhao, L. S. Yan, P. Gao, C. T. Wang, N. Yao, K. P. Liu, B. Jiang, and X. G. Luo, Moiré fringes characterization of surface plasmon transmission and filtering in multi metal-dielectric films, Appl. Phys. Lett. **105**, 141107 (2014).

[7]     S. Balci, A. Kocabas, C. Kocabas, and A. Aydinli, Localization of surface plasmon polaritons in hexagonal arrays of Moiré cavities, Appl. Phys. Lett. **98**, 031101 (2011).

[8]     A. Kocabas, S. S. Senlik, and A. Aydinli, Slowing Down Surface Plasmons on a Moiré Surface, Phys. Rev. Lett. **102**, 063901 (2009).

[9]     S. S. Senlik, A. Kocabas, and A. Aydinli, Grating based plasmonic band gap cavities, Opt. Express **17**, 15541 (2009).

[10]    S. Balci, M. Karabiyik, A. Kocabas, C. Kocabas, and A. Aydinli, Coupled Plasmonic Cavities on Moire Surfaces, Plasmonics **5**, 429 (2010).

[11]    E. Karademir, S. Balci, C. Kocabas, and A. Aydinli, Lasing in a Slow Plasmon Moiré Cavity, ACS Photon. **2**, 805 (2015).

[12]    D. M. Koller, U. Hohenester, A. Hohenau, H. Ditlbacher, F. Reil, N. Galler, F. R. Aussenegg, A. Leitner, A. Trügler, and J. R. Krenn, Superresolution Moiré Mapping of Particle Plasmon Modes, Phys. Rev. Lett. **104**, 143901 (2010).

[13]    O. Aftenieva, M. Schnepf, B. Mehlhorn, and T. A. F. König, Tunable Circular Dichroism by Photoluminescent Moiré Gratings, Adv. Opt. Mater. **9**, 2001280 (2021).

[14]    See Supplemental Material starting page 15 of the present document for the PbS NC synthesis, the sample fabrication workflow, a full description of the experimental setup, an experimental determination of n_SP(λ), a sketch illustrating the formation of the moiré fringes and two supplementary figures further illustrating the behavior of the model.

[15]    I. Moreels et al., Size-Dependent Optical Properties of Colloidal PbS Quantum Dots, ACS Nano **3**, 3023 (2009).

[16]  D. Schanne, S. Suffit, P. Filloux, E. Lhuillier, and A. Degiron, Spontaneous Emission of Vector Vortex Beams, Phys. Rev. Appl. **14**, 064077 (2020).

[17]  D. Schanne, S. Suffit, P. Filloux, E. Lhuillier, and A. Degiron, Shaping the spontaneous emission of extended incoherent sources into composite radial vector beams, Appl. Phys. Lett. **119**, 181105 (2021).

[18]  W. L. Barnes, T. W. Preist, S. C. Kitson, and J. R. Sambles, Physical origin of photonic energy gaps in the propagation of surface plasmons on gratings, Phys. Rev. B **54**, 6227 (1996).

[19]  In addition, if the periodicity is $P$ (as opposed to, e.g., $P/2$), the duty cycle of the $E(y)$ function must be different from 50%., For example, a $F(y)$ square function of periodicity $P$ and duty cycle $d$ also generates a moiré pattern (as long as d ≠ 50%), as also does a $P/2$-periodic square function.

[20]  B. Zhang, J. Zerubia, and J.-C. Olivo-Marin, Gaussian approximations of fluorescence microscope point-spread function models, Appl. Opt. **46**, 1819 (2007).

[21]  R. Röhrich and A. F. Koenderink, Double moiré localized plasmon structured illumination microscopy, Nanophoton. **10**, 1107 (2021).

# Supplemental Material for

# Plasmonic interferences sampled by a diffraction grating generate moiré fringes

Marin Tharrault[1], Guillaume Boulliard[1], Emmanuel Lhuillier[2], Aloyse Degiron[1,*]

[1]Université Paris Cité, CNRS, Laboratoire Matériaux et Phénomènes Quantiques, 75013 Paris, France

[2]Sorbonne Université, CNRS, Institut des NanoSciences de Paris, INSP, 75005 Paris, France

## 1. Colloidal PbS nanocrystal (NCs) synthesis

100 µL of trioctylphosphine (Alfa Aesar, 90%) along with 300 mg of $PbCl_2$ (Alfa Aesar, 99%) and 7.5 mL of oleylamine (OLA, Fisher Scientific, 90%) are first degassed at room temperature and then at 110 °C in a three-neck flask for 30 min. Meanwhile, 30 mg of sulfur powder (Alfa Aesar, 99.5%) are mixed with 7.5 mL of OLA until full dissolution. The atmosphere of the three neck flask is switched to nitrogen and the temperature cooled to 80 °C. This clear orange-looking sulfide solution is quickly injected into the three neck flask. We stop the reaction after two min by adding a mixture made of 1 mL of OA and 9 mL of hexane. The temperature is further cooled by flowing fresh air onto the three neck flask surface. We then precipitate the nanocrystal (NCs) with ethanol (VWR, >99.9%) and then redisperse them in 3 mL of toluene (VWR, rectapur). After repeating this washing step a second time, we disperse the pellet in 5 mL of toluene with a drop of OA. The supernatant is precipitated with methanol and redispersed in toluene. The solution is then centrifugated without addition of non-solvent to eliminate the colloidally unstable phase. This solution of PbS NCs in toluene is finally filtered through a 0.2 µm PTFE filter.

## 2. Sample fabrication

200 nm of Au are coated on a Si substrate using a Plassys MEB 550S electron-beam evaporator. The gratings are subsequently defined by applying a layer of CSAR62 resist from Allresist onto the sample and patterning this resist with an electron beam lithography system (Raith Pioneer II). 50 nm of Au are then evaporated on the sample and the remaining resist is lifted-off. Finally, a single layer of PbS NCs is spin-coated on the sample. This layer is approximately 20 nm thick.

The choice of a 1.35 µm grating period allows the diffraction of the surface plasmon modes within the 0.3 numerical aperture of our microscope objective [see the dispersion relation Fig. 2b]. The choice of a 50 nm high grating results from a tradeoff between a decreased intensity of the diffracted signal and a reduced surface plasmon propagation distance.

## 3. Experimental details

All experiments are carried on with a BX51WI microscope from Evident (formerly known as Olympus).

*Photoluminescence spectra*

The photoluminescence spectrum of Fig. 1(c) is measured on a flat area of the sample (i.e. without grating), using a HeNe laser to excite the NCs through a 10X objective. This objective is also used to collect the photoluminescence, which is subsequently filtered from the pump using a dichroic mirror (Thorlabs ref. DMLP950R) and a longpass filter (Thorlabs ref. RG780). The signal is focused on the front slit of an imaging spectrometer composed of an Acton SP2356 imaging monochromator coupled to a NIRvana 640ST InGaAs camera from Teledyne Princeton Instruments.

*Dispersion relations*

To measure the dispersion of Fig. 2(b), we start from the experimental configuration depicted in Fig. 1(a), producing the image displayed in Fig. 2(a), and insert an additional Bertrand lens in the optical path so as to image the back focal plane of the objective (Fourier imaging). This image is spatially filtered by the entrance slit of our imaging spectrometer and finally dispersed as a function of the wavelength by the Acton monochromator coupled to the InGaAs camera. Figure S1(a) displays the raw data, which show that the plasmonic branches are accompanied by an isotropic photoluminescence background. This background light corresponds to the fraction of the photoluminescence that is not coupled to the surface plasmons. Figure S1(b) shows that this background light is present when we repeat the experiment on a flat area of the sample devoid of metallic grating, which makes us conclude that it originates from the NCs that are located directly under the laser excitation spot. Since we are only interested by the light emitted by the grating, the data displayed in the main text as Fig. 2(b) and reproduced here in Fig. S1(c) have been corrected using Fig. S1(b) to suppress the isotropic background.

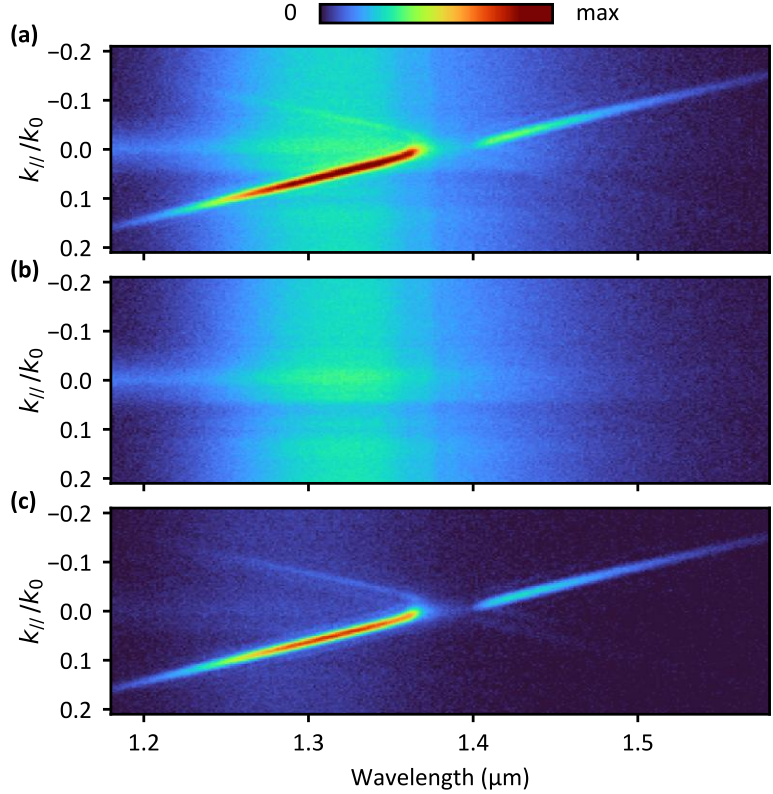

FIG. S1. (a) Raw dispersion relation, with light originating both from the uncoupled photoluminescence of the NCs excited under the laser pump and from the plasmonic grating $I_{mes}$. (b) Same experiment, but performed in a region of the sample devoid of plasmonic grating $I_{bkd}$. (c) Corrected dispersion relation, computed from the relation $I_{cor} = I_{mes} - 0.8\, I_{bkd}$. The 0.8 coefficient accounts for the fact that the nearly-isotropic photoluminescence of panel b is higher than in panel a.

*Experimental moiré patterns*

We also start from the experimental configuration depicted in Fig. 1(a), producing the image displayed in Fig. 2(a). We form this image at the entrance slit of our imaging spectrometer. We use the slit to select the vertical slice of this image that crosses the x=0 abscissa. This slice is then horizontally dispersed as a function of the wavelength by the Acton monochromator coupled to the InGaAs camera. In other words, the only difference compared to the experimental configuration used to measure the dispersion relations is that no Bertrand lens is inserted in the optical path, so that the vertical axis is position space instead of momentum space. A black adhesive, glued in the imaging plane of the spectrometer slit, is used to mask the strong photoluminescence below the illumination spot, while allowing to collect the light scattered by the grating 100 µm away. Additionally, a small imaging tilt aberration from the imaging system – arising from the large field-of-view – is numerically corrected to produce the moiré patterns displayed in this study.

**4. Experimental determination of the complex effective index of the surface plasmons**

The wavelength-dependent effective index, defined as $n_{SP}(\lambda) = k_{SP}/k_0$, can be extracted from dispersion measurements. To this end, we measure the dispersion of surface plasmons excited at the center of a 180-µm long Au grating. The result is plotted in Fig. S2(a). Contrarily to the dispersion plotted in the main text as Fig. 2(b), the plasmonic branches are symmetric because we focus the laser pump at the center of the grating, which means that the plasmon propagation away from this excitation spot is also symmetrical. The real part of $n_{SP}(\lambda)$ is obtained from the positions of the plasmonic branches, by virtue of the momentum conservation law that can be rewritten as $Re(n_{SP}) = k_{//}/k_0 + G/k_0$, while its imaginary part can be derived from the width of these same branches. Strictly speaking, the branch width $\Delta\theta$ provides a measurement of the spatial coherence length of the plasmons, which is equal to $\lambda/\Delta\theta$ for an infinitely long grating. Due to the absorption of Au and the high radiative losses induced by the grating, this spatial coherence length is also the characteristic propagation length of the plasmons, also given by $1/[2Im(n_{SP})k_0]$. The real part of $n_{SP}$ and the plasmon propagation length are plotted in Fig. S2(b) and S2(c), respectively.

We did not directly extract $n_{SP}(\lambda)$ from the dispersion relation plotted in Fig. 2(b) because it has been measured with a grating having a lateral extension of only 80 µm. In other word, the grating considered in Fig. 2(b) cannot be considered as infinitely long, preventing us from using the branch width $\Delta\theta$ to retrieve the propagation length.

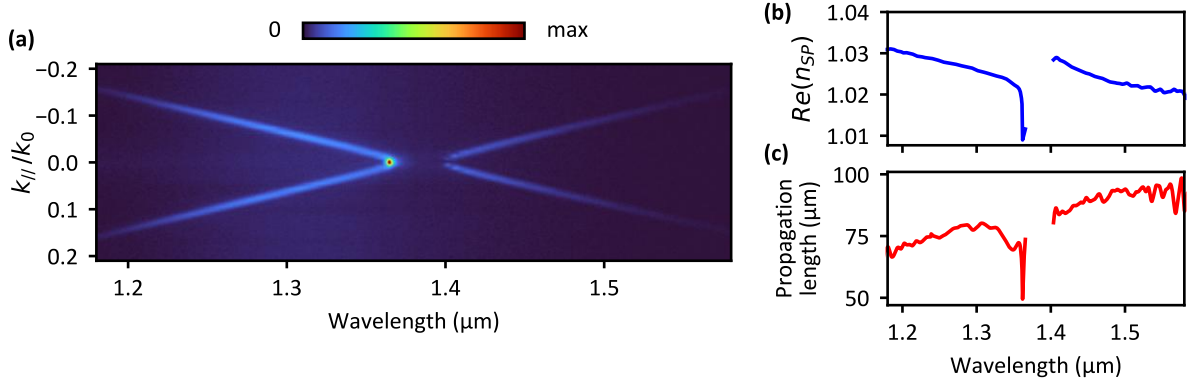

FIG. S2. (a) Dispersion relation of surface plasmons launched at the center of a 180 µm long grating, with the same groove height, duty cycle and periodicity as the other gratings considered in this study. (b) Extracted real part of the effective index Re($n_{SP}$). (c) Extracted propagation length.

## 5. Supplementary figures

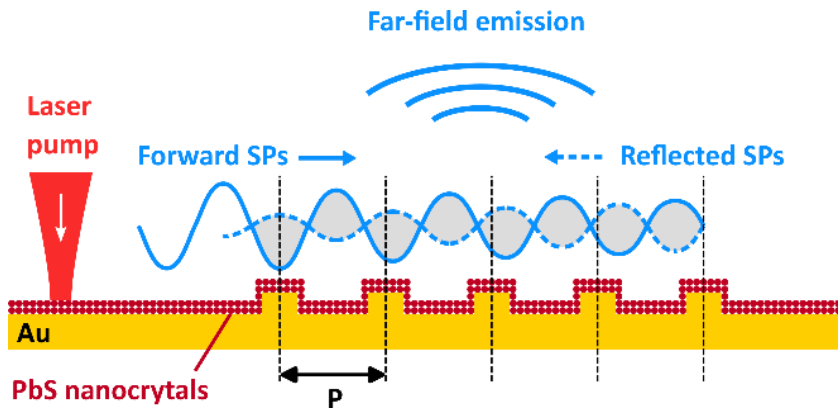

FIG. S3. Schematic of the moiré generation mechanism. The active PbS nanocrystal layer is pumped by a red laser, decaying in surface plasmons that freely propagate over the Au surface to the grating. As the grating has a finite extent, it results in the formation of forward and reflected plasmons that are scattered to the far-field by the grating. The interference of these waves forms a standing wave pattern that is sampled with a P periodicity, giving rise to a characteristic moiré pattern in wavelength – position parameter space.

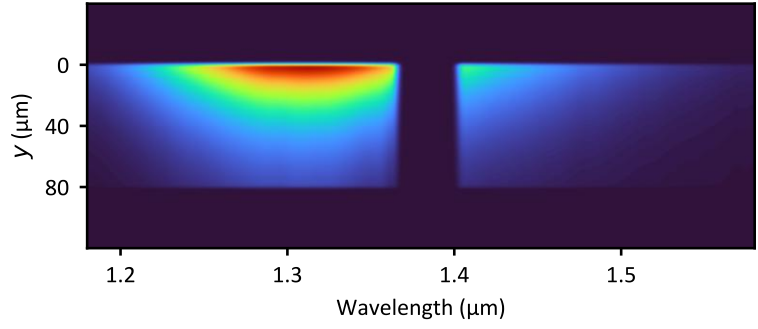

FIG. S4. Computed evolution of $|A(\lambda, y)|^2$ convoluted with the point spread function of our apparatus to emulate the experimental resolution. The moiré pattern is not present due to the absence of the P-periodic scattering mechanism.

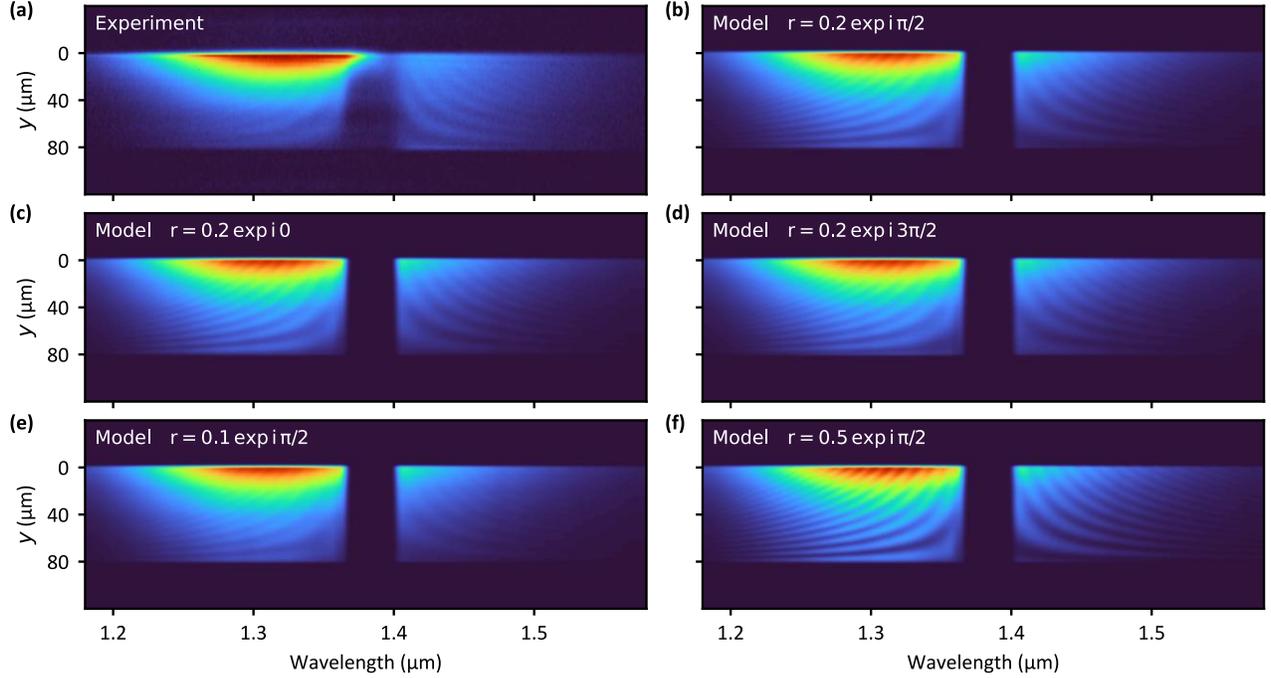

FIG. S5. Effect of the $r$ reflection coefficient on the moiré pattern. (a) Experimental moiré pattern for the 80 µm-long grating, as displayed in Fig. 2(c) and Fig. 3(b). (b-f) Computed moiré patterns for different complex coefficients $r$. The case that replicates the experimental data in the most faithful way is panel b, corresponding to a reflection coefficient $r = 0.2 \exp(i\pi/2)$. The calculations displayed in the main text use this specific coefficient.